\documentclass[10pt]{iopart}
\usepackage{graphicx}

\begin{document}

\title{Tunable sub-gap radiation detection with superconducting resonators}


\author{O.~Dupr\'e $^{1,2}$ A.~Benoit $^{1,2}$, M.~Calvo $^{1,2}$, A.~Catalano $^{1,2, 3}$, J.~Goupy $^{1,2}$, C.~Hoarau $^{1,2}$, T.~Klein $^{1,2}$, K. Le Calvez $^{1,2}$, B.~Sac\'ep\'e $^{1,2}$, A.~Monfardini $^{1,2}$, F.~Levy-Bertrand $^{1,2}$.}
\address{$^1$ Univ. Grenoble Alpes, Inst NEEL, F-38000 Grenoble, France}
\address{$^2$ CNRS, Inst NEEL, F-38000 Grenoble, France}
\address{$^3$ Laboratoire de Physique Subatomique et de Cosmologie, Universit\'e Grenoble Alpes, CNRS/IN2P3, 53 rue des Martyrs, 38026 Grenoble Cedex, France}
\eads{alessandro.monfardini@neel.cnrs.fr}
\eads{florence.levy-bertrand@neel.cnrs.fr}

\date{\today}

\begin{abstract}
We have fabricated planar amorphous Indium Oxide superconducting resonators ($T_c\sim2.8$~K) that are sensitive to frequency-selective radiation in the range of 7 to 10~GHz. Those values lay far below twice the superconducting gap that worths about 200~GHz. The photons detection consists in a shift of the fundamental resonance frequency. We show that the detected frequency can be adjusted by modulating the total length of the superconducting resonator. We attribute those observations to the excitation of higher-order resonance modes. The coupling between the fundamental lumped and the higher order distributed resonance is due to the kinetic inductance non-linearity with current. These devices, that we have called Sub-gap Kinetic Inductance Detectors (SKIDs), are to be distinguished from the standard Kinetic Inductance Detectors (KIDs) in which quasi-particles are generated when incident light breaks down Cooper pairs.
\end{abstract}

%
\vspace{2pc}

\noindent{To cite: O. Dupre et al, Supercond. Sci. Technol.  30, 04007 (2017).}
%
%
\ioptwocol

\section{Introduction}
Superconducting microwaves resonators are the building blocks of various ongoing and future technological developments ranging from sensitive photon detectors for astrophysics~\cite{NIKA}, superconducting quantum devices~\cite{Wallraff}, and coupling to nano-electromechanical resonators~\cite{Regal}. Kinetic Inductance Detectors (KIDs)~\cite{Day} are a particular implementation of superconducting resonators. The detection principle is based on monitoring variations of the resonator frequency $\omega_0=(LC)^{-1/2}$. The incident radiation breaks down Cooper pairs, modifying the kinetic inductance  $L_K$ and thus the resonance frequency  ($L=L_K+L_{geometrical}$). 

The resonator superconducting material directly affects its employability. Indeed, the critical superconducting temperature $T_c$ imposes a working temperature $T<<T_c$ and the superconducting gap $\Delta$ sets in the photon detector cutoff frequency to $h\nu>2\Delta$. Within the BCS-superconducting theory  $T_c$ and $\Delta$ are not independent parameters as  $\Delta~=~1.76\times K_BT_c$. Thus, for classic KID lowering the cutoff frequency requires to lower accordingly the operating temperature. For example to detect 100~GHz one needs an operating temperature $T\sim100~$mK, and to detect 10~GHz  $T\sim10~$mK is required. 

In this letter, we present a disruptive technology  for millimetric down to centimetric detection that we have called Sub-gap Kinetic Inductance Detectors (SKIDs). These detectors are sensitive to photons with an energy $h\nu$ laying well below twice the superconducting gap $2\Delta$ thus removing the operating temperature constraint when lowering the photon detection cutoff frequency. We attribute their detection mechanism to the excitation of higher-order resonance modes combined with the kinetic inductance non-linearity with current. In case of disordered superconductors, the kinetic inductance non-linearity with current is more pronounced because of a reduced critical current due to a low superfluid density~\cite{Lorentz}. Thus, disordered superconductors are of interest to develop sensitive SKIDs.

In that context, we have investigated amorphous Indium Oxide films, $a:InO$, that can be tuned through the superconductor-to-insulator transition by varying the disorder level~\cite{InO_tunning1, InO_tunning2}. Lumped superconducting resonators have been designed and fabricated. The resonators fundamental frequencies variations have been monitored while illuminating at higher frequency. At centimetric wavelengths we studied the power illumination influence to evaluate the detectors sensitivity order of magnitude. We discuss our observations in term of  higher-order distributed resonance modes excitations and kinetic inductance non-linearity with current mechanism.

\section{Methods}

\begin{figure}[t]
\begin{center}
\resizebox{8.5cm}{!}{\includegraphics{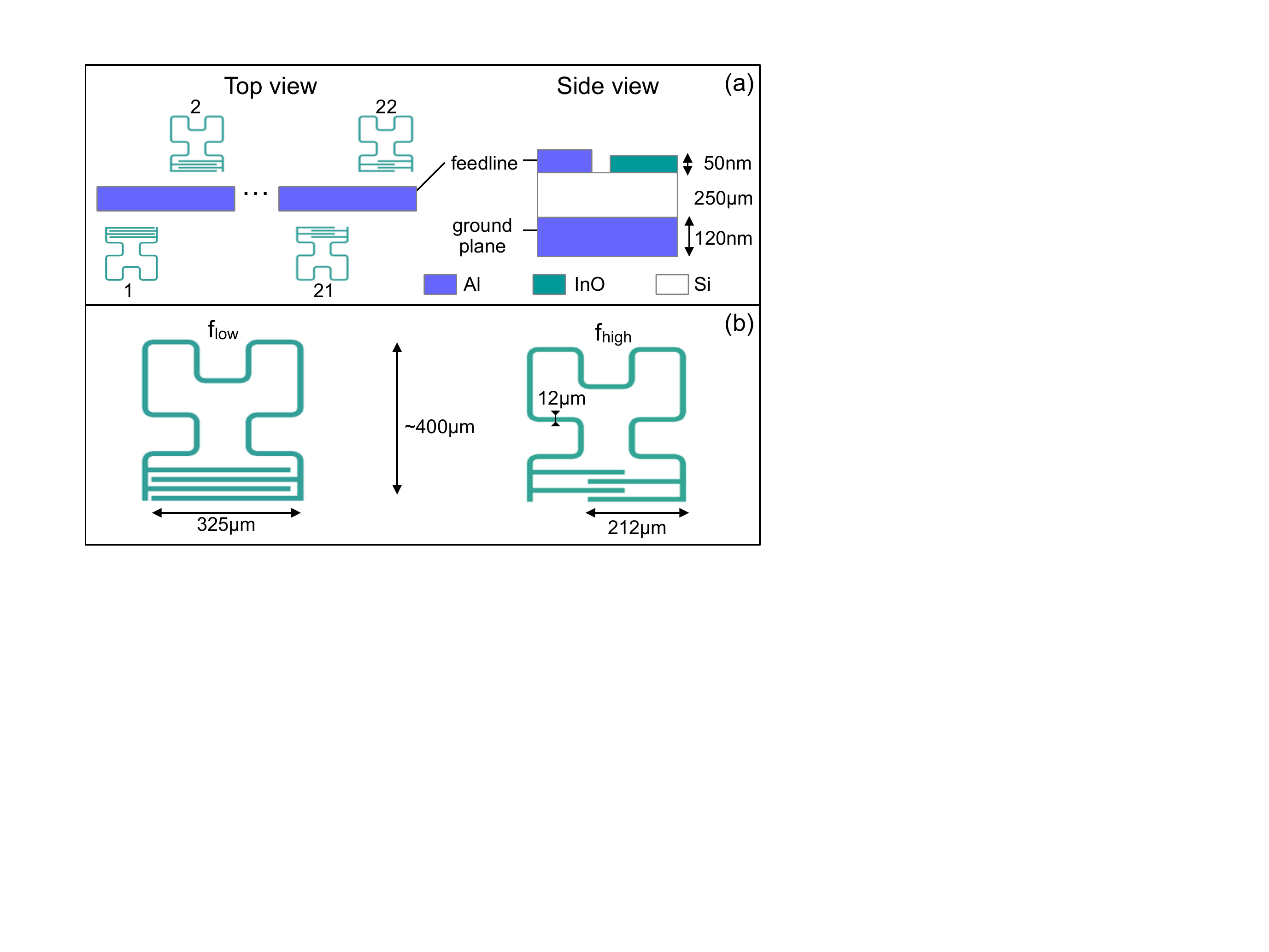}}
\caption{\textbf{ Resonators design.} (a) Twenty-two devices are coupled to the common feed-line. Cross-section view of the micro-strip configuration. (b) Detailed design of the $f_{low}$ and the $f_{high}$ resonators. The Hilbert fractal inductor is identical for all the resonators. The four capacitor fingers length are varied from one resonator to another: from the shortest ($f_{high}$) to the longest ($f_{low}$). }
\label{design}
\end{center}
\end{figure}

Figure~\ref{design} displays the resonators design. Twenty-two $a:InO$ resonators were deposited on a silicon substrate. They are coupled to a $50\Omega$ aluminum micro-strip transmission line. Each resonator consists of a second order Hilbert shape fractal inductor and an interdigitated capacitor\cite{Hibert,Roesch}. Frequency multiplexing is achieved by varying the capacitor fingers length. The design of the lowest ($f_{low}$)  and the highest frequency ($f_{high}$) resonators are detailed on panel (b). The  $f_{low}$ and $f_{high}$ resonators correspond to a maximum wire total length of respectively 2.3~mm and 2.1~mm. In relatively dark conditions we estimate the average internal quality factor to be of the order of 28000  (ranging from 23000  up to  36000).  A four wires resistivity measurement of the 50~nm $a:InO$ film as a function of temperature allows to determine the critical superconducting transition temperature $T_c$ and the sheet resistance $R_s$: $T_c\sim2.8~K$ (defined at the foot of the transition) and $R_s(T=5~K)~=~1870~\Omega/sq$. Tunneling measurements established that twice the superconducting gap of the employed Indium Oxide worths about 200~GHz, see supplementary Information in reference~\cite{Sacepe}. Using the BCS theory in the dirty limit at low frequency $f$ and temperature $T$, i.e $hf<<\Delta$ and  $K_BT<<\Delta$, we can estimate the sheet inductance $L_s$ from the following formula~\cite{Rotzinger} $L_s\sim0.18\times(\hbar R_s)/(K_BT_c)$ to be of the order of $\sim$900~pH/sq (assuming that $\Delta=1.76K_BT_c$). 

\section{Results}

\begin{figure}[t]
\begin{center}
\resizebox{5.5cm}{!}{\includegraphics{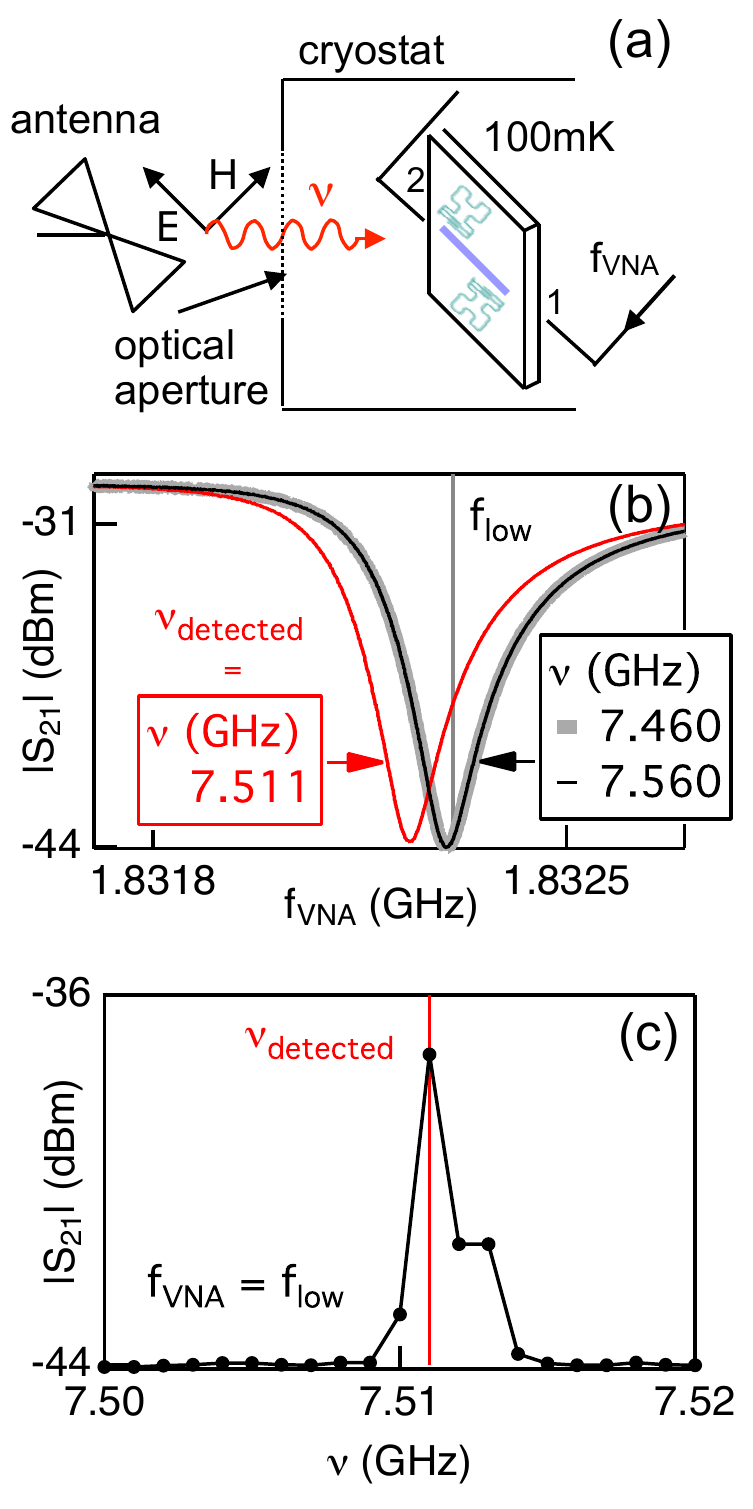}}
\caption{\textbf{GHz sub-gap detection demonstration.} (a) Experimental set-up. A bow-tie antenna at room temperature illuminates at a $\nu$-frequency the resonators through the optical apertures of the dilution refrigerator.  (b) $f_{VNA}$-scans for different fixed $\nu$-antenna frequencies. Vertical line: $f_{low}$ is the unperturbed resonance frequency. Only $\nu_{detected}$~=~7.511~GHz gives a measurable shift of the resonance. (c) $\nu$-scan at a fixed VNA-frequency $f_{VNA}=f_{low}$. The peak center at $\nu_{detected}$~=~7.511~GHz corresponds to the resonance frequency shift presented on panel (b).}
\label{first_results} 
\end{center}
\end{figure}

 In figure~\ref{first_results} we present the main result of this paper. Panel (a) shows a sketch of the experimental set-up. The resonators were cooled down at 100~mK with a $^3$He/$^4$He dilution refrigerator. A custom bow-tie antenna~\cite{antenne} was placed at room temperature in front of the optical aperture of the cryostat to  illuminate the resonators in the GHz-range.  The bow-tie antenna is composed by two identical planar metallic triangles that are bilaterally symmetrical. The triangles form a dipole that irradiates mainly perpendicular to its plane, over a broad frequency band. The length of the antenna sets the frequency bandwidth~\cite{antenne} that is in this case from 4~GHz up to 15~GHz. A radio-frequency generator was connected to the antenna and allowed to vary the illumination frequency up to 15~GHz. The light emitted from the antenna will be further referred as $\nu$. The resonators frequencies were monitored employing a Vectorial Network Analyzer (VNA) connected to the transmission line. Using the VNA we measured the $S_{21}$ transmission coefficient for two types of scans : $f_{VNA}$-scans and $\nu$-scans. In $f_{VNA}$-scans we spanned the VNA-frequency for a fixed $\nu$-antenna frequency. In $\nu$-scans we varied the $\nu$-antenna frequency while measuring the resonator $S_{21}$ transmission coefficient  at a fixed $f_{VNA}$ frequency. Panels (b) and (c) detail the results obtained on the resonator with the lowest resonance frequency $f_{low}\sim1.832$~GHz. Panel (b) shows three $f_{VNA}$-scans performed at different $\nu$-antenna frequencies. The unperturbed resonance frequency, $f_{low}$, is unaffected by all the $\nu$-antenna frequencies but $\nu_{detected}$~=~7.511$\pm$0.003~GHz. For  $\nu_{detected}$ the resonance is shifted to lower frequencies. The peak in panel (c) corresponds to the resonance frequency shift observed on panel (b) and indicates the $\nu$-antenna detected by the resonator. In summary, within the whole range 7-10~GHz, the $f_{low}$-resonator detects only a radiation peaked at 7.511~GHz with a frequency selectivity of $\pm$0.003~GHz. The resulting resolving power is $R=\nu/d\nu=2500$.

 \begin{figure}[t]
\begin{center}
\resizebox{5.5cm}{!}{\includegraphics{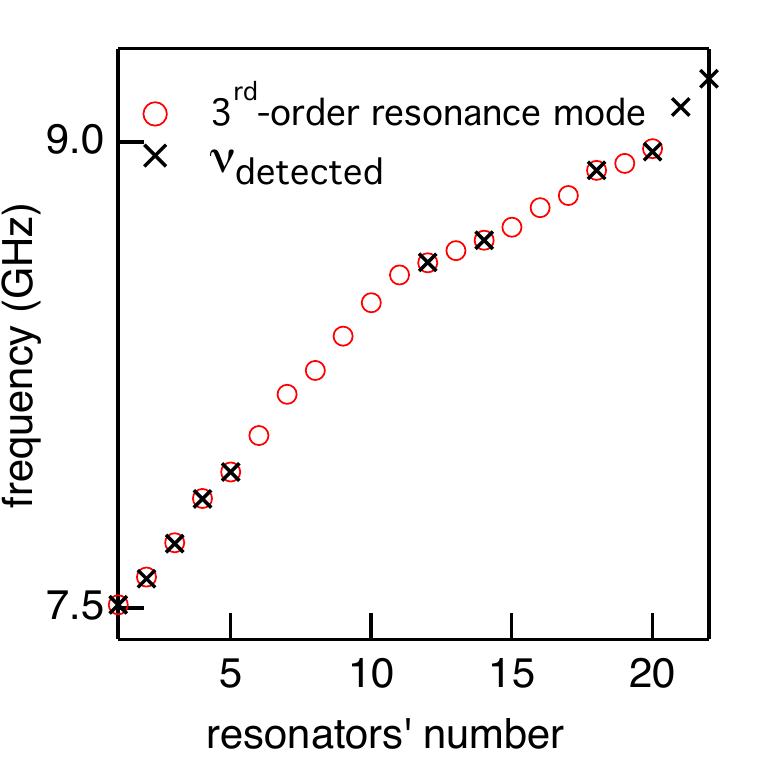}}
\caption{\textbf{GHz sub-gap detection tunability.} As a function of the resonator number the $\nu_{detected}$-frequency compared to third-order resonance mode identified with a $f_{VNA}$-scan. From one resonator to the next the total length varies by $\sim10\mu$m. The $f_{VNA}$-scan was measured up to 9~GHz.  The $\nu_{detected}$-frequency was probed for eleven resonators.}
\label{concept_tunabilty}
\end{center}
\end{figure}

Figure~\ref{concept_tunabilty} compares, as a function of the resonator number, the $\nu_{detected}$-frequency for eleven resonators to the third-order resonance mode measured with a $f_{VNA}$-scan up to 9~GHz. When measured, the $\nu_{detected}$-frequency  equals the third-order resonance mode. The fundamental resonance frequencies of the resonator number 1 and number 22 are respectively $f_{low}~\sim$~1.832~GHz and $f_{high}~\sim$~2.552~GHz. From one resonator to the next the total length changes by $\sim10\mu$m. Within the 7-10~GHz range, each resonator detects only a $\nu_{detected}$-frequency with a resolving power R of several thousands (like the $f_{low}$ shown in figure~\ref{first_results}c). The $\nu_{detected}$-frequency is different for each resonator, varying  from 7.511~GHz for the first resonator to  9.204~GHz for the last resonator. The detected frequency corresponds to the third-order resonance mode measured with a $f_{VNA}$-scan.  Being a distributed resonance, the detected frequency varies with the total length of the superconducting resonator, and, thus, can be adjusted by modulating the total length of the resonator. The inductor, the radiation-sensitive structure, is identical for all the resonators.

\begin{figure}[t]
\begin{center}
\resizebox{8.5cm}{!}{\includegraphics{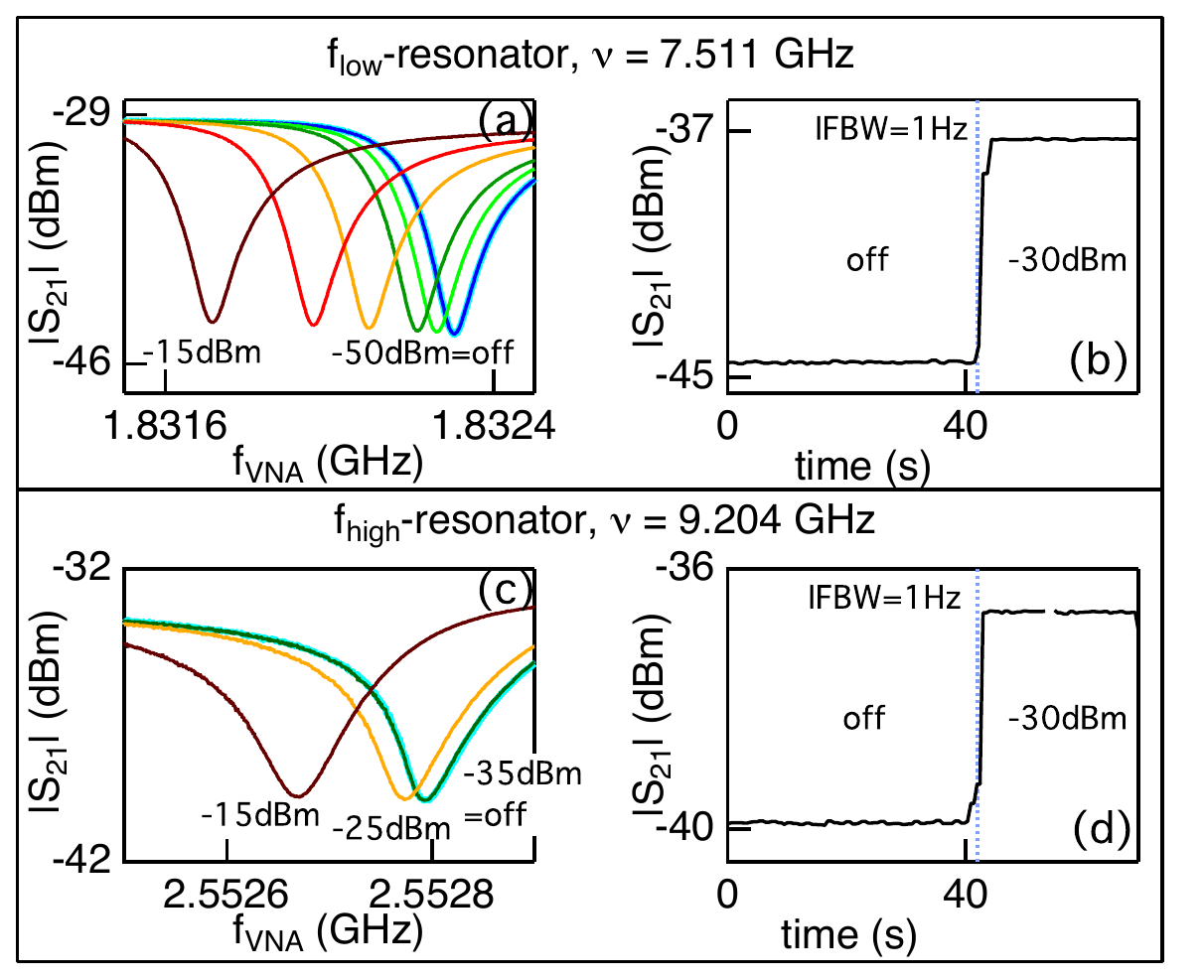}}
\caption{\textbf{Power and noise  studies.} Left: $f_{VNA}$-scans for different $\nu$-antenna power. Power applied on the antenna is, from right to left, in dBm: (a) off, -50, -45, -35, -25,- 20, -15 (c) off, -35, -25, -15. Right: time variation of $|S_{21}|$ measured at the unperturbed resonator frequency with an integrated bandwidth frequency of 1~Hz. The $\nu$-antenna is off until $t\sim$40~s and on after, with -30~dBm applied on the antenna. (b)~$|\Delta S_{21}|$~=~7.22~$\pm$~0.03~dBm (d)~$|\Delta S_{21}|$~=~3.24~$\pm$~0.02~dBm.}
\label{power_NEP}
\end{center}
\end{figure}

Figure~\ref{power_NEP} presents power and noise studies conducted on the $f_{low}$ and $f_{high}$ resonators to evaluate the order of magnitude of their sensitivity. On the left, the  power study revealed that the frequency shift due to the $\nu$-antenna frequency increases with the incident power. The indicated power corresponds to the one applied onto the antenna with the RF-source. On the right, the noise study consists in registering the time variation of $|S_{21}|$ measured at the unperturbed resonator frequency with an integrated bandwidth frequency of 1~Hz while the  $\nu$-antenna is off and switched on at $t\sim$40~s with -30~dBm applied onto the antenna. The standard deviation without illumination is (b)  rms~=~0.03~dBm and (d) rms~=~0.02~dBm. The signal variation is (b) $|\Delta S_{21}|$~=~7.22~dBm and (d) $|\Delta S_{21}|$~=~3.24~dBm. From these results we can approximately estimate the sensitivity of our Sub-gap Kinetic Inductance Detectors.  We define $P_{SKID}$ as the power hitting each resonator. Electromagnetic simulation of the antenna gives at 8~GHz a radiation efficiency of 75$\%$ and a directivity of 2.7dBi. In the case of $P_{antenna}~=~-30~dBm$, three dimensional simulation ray-tracing with the cryostat geometry roughly estimates  $P_{SKID}\sim10^{-12}$~W. The Noise Equivalent Power (NEP) is defined as:
\begin{equation}
NEP=P_{SKID}/\frac{signal}{noise}
\label{NEP}
\end{equation}
with the noise measured for a 1~Hz bandwidth. The signal to noise ratio is of the order of $\sim$200$\sqrt{Hz}$. Thus, assuming for simplicity a linear response of the detector (implicit in the above equation), the rough evaluation of the NEP is in the $10^{-14}-10^{-15}$~W/$\sqrt{Hz}$ range. For future development it is important to notice that the detector response is actually nonlinear : see figure~2 of the supplementary material.

Radio-frequency electromagnetic simulations have been realized using the Sonnet software~\cite{SONNET}. We adjusted the kinetic inductance $L_s$~=~700~pH/sq to obtain the same $f_{low}$-resonance frequency by simulation and measurement (1.83~GHz). With the adjusted $L_s$ value, we get for the  $f_{high}$-resonance a simulated resonance frequency of 2.57~GHz which corresponds  within 1\% to the 2.55~GHz measured frequency. For the $f_{low}$-geometry, besides the base resonance at 1.83~GHz, the simulation revealed higher frequency resonance modes at  6.46~GHz and 7.53~GHz.  For the $f_{high}$-geometry, besides the base resonance at 2.57~GHz, higher frequency resonance modes are found at  6.81~GHz and 9.16~GHz. The highest simulated frequencies match within less than 1\% the $\nu$-frequencies detected.  Within the present set-up our resonators did not detect the $\sim$~6~GHz-frequency but they did with an antenna enclosed in the cryostat and placed at 1~cm from the resonators. This probably indicates that the current induced from the incident radiation in the resonator or the cross-kerr coupling between the fundamental and the harmonic mode is weaker for the second resonance order compared to the third. The induced current is determined by the optical absorption efficiency, the internal and the coupling quality factors of the resonator, the cross-kerr coupling quantifies the frequency shift of the fundamental mode when an harmonic mode is excited. We have also detected $\nu$-frequencies above the 7-10~GHz range. The observations were similar: the $\nu$-detected frequency was different for each resonator with a very high frequency-selectivity. Again, the $\nu$-frequencies detected correspond to simulated frequencies. The agreement between the simulated and/or the measured higher-order resonance mode and the detected frequencies lead us to conclude that the higher-order resonance modes are at the origin of the detection mechanism of our SKIDs. In the following paragraph we discuss why the fundamental resonance mode is shifted when a harmonic mode is excited.  

ÒThere is a $\sim25\%$ difference from the $\sim$900pH/sq estimated with the BCS-analytic formula and the 700pH/sq extracted from the comparison between simulated and measured resonance frequencies. The difference is likely due to the non BCS-behavior of a:InO [14]. For small penetration depth compared to the resonator width, another source of difference may be due to the fact that the electromagnetic simulation assumes an uniform current distribution for each step grid throughout the width of the resonator. This last point is inaccurate. Adjusting by simulation the kinetic inductance per square to match the experimental frequencies might thus lead to an underestimation of the kinetic inductance. This source of discrepancy is probably minor here as for a step grid of $3~\mu m$ almost no change of the current distribution was observed throughout the $12~\mu m$-width. Finally, from a practical point of view, the effective kinetic inductance extracted this way is the most useful value to foresee the highest harmonic modes.Ó

\section{Discussion}

Below twice the superconducting gap, in order to be absorbed, the frequency of an incident light radiation has to match a resonance mode (or any collective mode) otherwise the superconductor is a perfect mirror. When absorbed, the radiation increases the resonator current density, modifying the inductance $L$ owing to the following expression~\cite{Lorentz}:
\begin{equation}
L(J)=L(0)[1+J^2/J_*^2+..]
\label{inductance_nonlinearity}
\end{equation}
where $J_*$ is a constant that sets the scale of the kinetic inductance nonlinearity. The change of inductance leads to a shift of the fundamental resonance frequency.

Within the Ginsburg-Landau theory~\cite{DeGennes} $J_*=2/3^{3/2}J_c$ where $J_c$ is the critical current density. We can estimate the current density $J_*$ using the following formula~\cite{DeGennes} $J_*=\hbar e /m \times n_s / \xi $ where $\hbar$ is the reduced Planck constant, $e$ is the elementary charge,  $m$ is the electron mass, $n_s$ is the superfluid density (at zero temperature), and $\xi$ is the Cooper pairs' coherence length (at a given temperature). 
For the employed $a:InO$ with $n_s\sim10^{24}~m^{-3}$ (see supplementary material) and $\xi=5~nm$ (from reference~\cite{Sacepe}) we get $J_*\sim~4\times10^{9}~A/m^{-2}$. This value is to be compared to the critical current density of granular aluminum~\cite{Buisson}  $J_*\sim~3\times10^{10}~A/m^{-2}$ or pure aluminum~\cite{Kittel} $J_*\sim~1\times10^{12}~A/m^{-2}$. 


Low superfluid density are of importance as it contributes to lower $J_*$ and thus increases the detector sensitivity~\cite{Lorentz}. Another way to improve the detector sensitivity is to enhance the current density $J$  by reducing the section of the resonator, by optimizing the coupling to the feed-line or by injecting a bias DC-current as realized in a frequency-tunable resonator~\cite{Vissers}. The DC bias would allow a dynamic adjustment of the operating point.

The non-linear behavior of the kinetic inductance is already adopted in the recently proposed kinetic inductance parametric amplifiers~\cite{Eom} and in a frequency-tunable resonator via a DC-current~\cite{Vissers}. Our developments are mainly motivated by the need of integrating low frequency channels into classical KID-based instruments. For example, in the proposed next generation satellite CORE, devoted to the ultimate study of CMB (Cosmic Microwave Background) polarization, the baseline focal plane is based on KID. CORE will nominally cover the band 60-600 GHz, split into at least 15 sub-bands in order to efficiently allow foreground components separation. Adding lower (e.g. 20-60 GHz) frequency channels would allow to strongly improve this procedure. The goal is to clearly identify the sources exhibiting non-thermal spectra, e.g. synchrotron radio emission. Using SKID instead of alternative technologies like bolometers or HEMT would allow using the existing readout electronics and naturally integrate the further pixels into the existing focal plane held at 100 mK. Further potential applications of the SKID include precise, real-time atmospheric monitoring for millimeter-wave interferometry~\cite{NOEMA} or multi-beam, intermediate energy resolution (R$\sim$1000) instrumentation on single-dish radio-telescopes.

\section{Conclusion}
In conclusion, we designed, fabricated and measured Kinetic Inductance Detectors that are sensitive to photons with an energy $h\nu$ laying well below twice the superconducting gap $2\Delta$ : the Sub-gap Kinetic Inductance Detectors (SKIDs). These detectors are innovative with respect to the current KID technology for two reasons: first  they remove the temperature constraint when lowering the photon detection cutoff frequency, second they are very selective in energy (resolution R$\sim$1000-10000). The frequency detected can be adjusted according to the total length of the resonator. The sub-gap detection mechanism is due to the excitation of higher-order resonance modes combined with the kinetic inductance non-linearity with current. As the  kinetic inductance non-linearity with low current is a key ingredient to get sensitive detectors we suggest that low superfluid density materials, like indium oxide, granular aluminum or niobium nitride are of interest for future SKIDs development.

\section*{Acknowledgments}
We acknowledge O.~Buisson for the numerous useful discussions. We thank I.~Pop and L.~Gr\"unhaupt for suggestions about the non linearity of kinetic inductance with current. This work is supported by the French National Research Agency through Grant No. ANR-12-JS04-0003-01 SUBRISSYME.

\vspace{1cm}
\bibliographystyle{unsrt}

\clearpage

\section*{Supplementary information}

\begin{figure}[h]
\begin{center}
\resizebox{8.5cm}{!}{\includegraphics{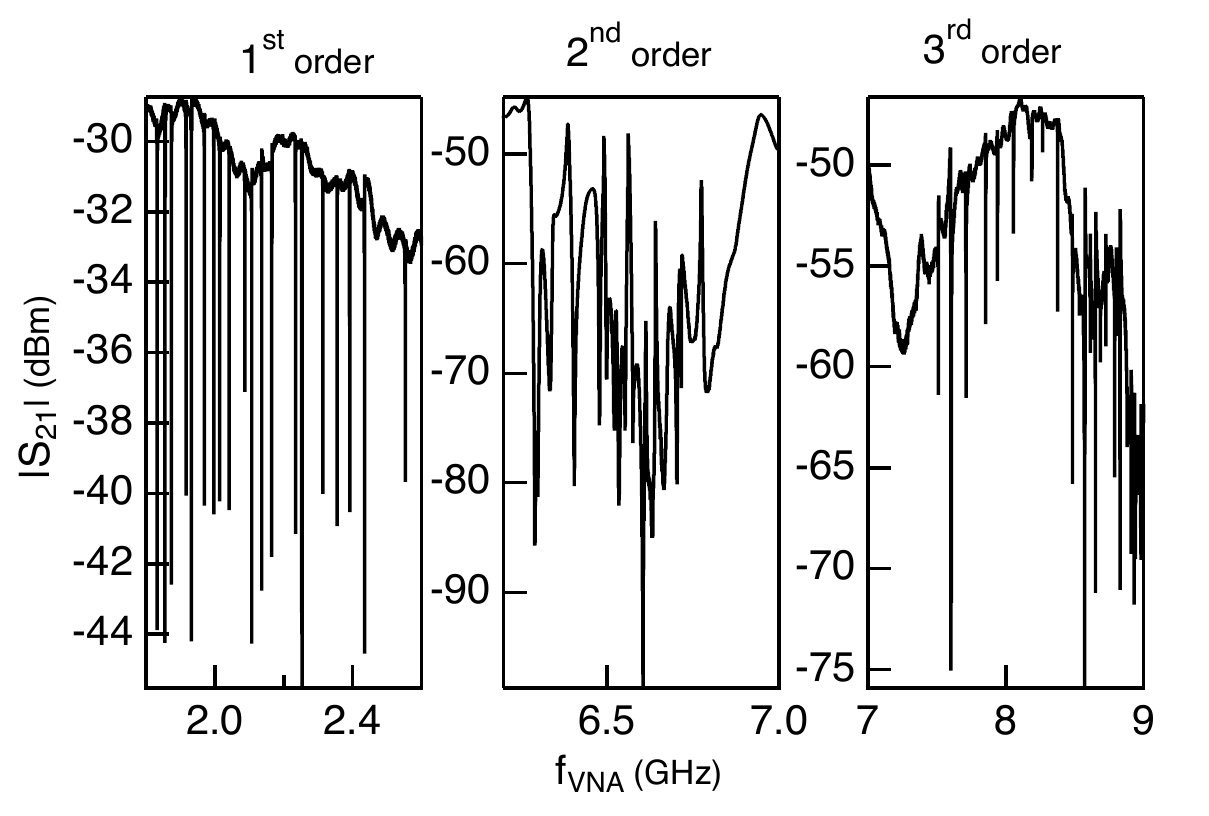}}
\caption{\textbf{$f_{VNA}$ scans up to 9GHz} Left: fondamental (lumped) resonances. Middle: second order resonances (first distributed modes). Right: third order resonances (second distributed modes).
}
\label{VNA_scans}
\end{center}
\end{figure}

Figure~\ref{VNA_scans} displays $f_{VNA}$ scans measured up to 9~GHz. The fondamental resonances were observed around 2~GHz (left panel). Very broad and overlapping second order resonances were observed around 6~GHz (middle panel). Third order resonances were found around 8~GHz (right panel).

The surface superfluid density $n_s$ of our $a:InO$ film is estimated from the sheet kinetic inductance $L_s$ using~\cite{Diener}:

\begin{equation}
n_s=\frac{m}{dL_se^2}
\label{ns_ls}
\end{equation}

\noindent where e is the elementary charge, $d~=~50~nm$ is the film thickness and $m$ is the electron mass. The $L_s$~=~700~pH/sq inductance has been adjusted to obtain the same $f_{low}$-resonance frequency by simulation and measurement (1.83~GHz). We get $n_s~=~1\times10^{24}$~m$^{-3}$.

\begin{figure}[h]
\begin{center}
\resizebox{5.5cm}{!}{\includegraphics{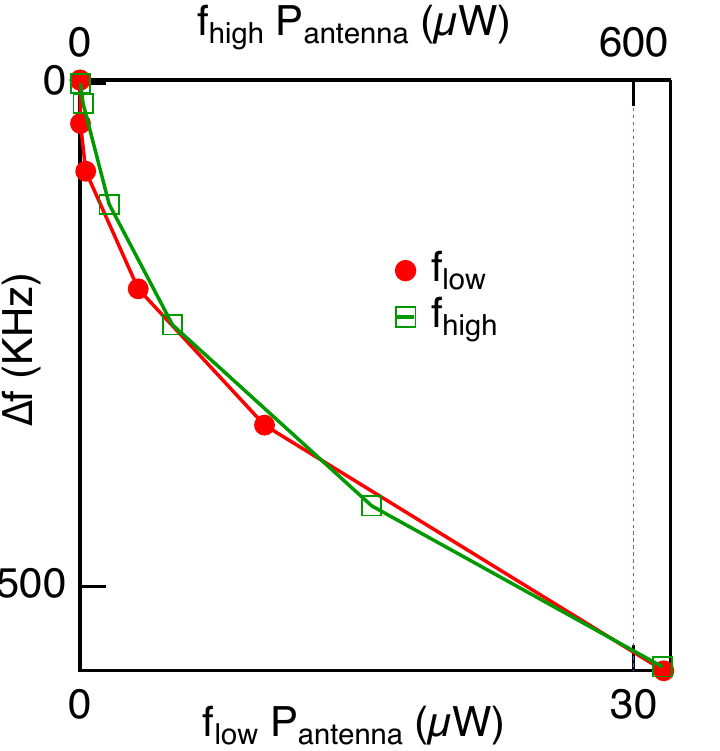}}
\caption{\textbf{Fundamental frequency shift as a function of the power applied on the antenna.} Bottom axis :  power applied on the antenna for $f_{low}$. Top axis : power applied on the antenna for $f_{high}$}
\label{deltaf}
\end{center}
\end{figure}

Figure~\ref{deltaf} shows the frequency shift of $f_{low}$ and $f_{high}$ as a function of the power applied on the antenna.  $f_{low}$ and $f_{high}$ are, respectively, the lowest
and the highest fundamental resonance frequencies. The inductor, the light-sensitive structure, is identical for all the resonators. The four capacitor fingers length are varied from one resonator to another: from the shortest ($f_{high}$) to the longest ($f_{low}$). The variation of the capacitance allows to achieve multiplexing and also modifies the coupling to the feed-line. Although 20 times more power is needed for $f_{high}$, the resulting shape of the frequency shift as a function of the linear power is identical. The kinetic inductance variation can be obtained almost directly from the frequency shift. On the contrary, the current density induced in the resonator by the antenna will depend, among other parameters, on the internal and coupling quality factors~\cite{Lorentz}. In conclusion, for a similar critical current density, the quality factors strongly influences the detector sensitivity. 

\bibliographystyle{unsrt}

 \end{document}